\documentclass[fleqn,usenatbib]{mnras}
\usepackage{newtxtext,newtxmath}
\usepackage[T1]{fontenc}
\usepackage{ae,aecompl}
\usepackage{braket}
\usepackage{bm}
\usepackage{graphicx}
\usepackage{color}
\usepackage{amsmath}
\usepackage{amssymb}
\usepackage{ulem}

\title[Inhomogeneous magnetic reheating]
{Secondary CMB temperature anisotropies from magnetic reheating}

\author[S.Saga et al.]{
Shohei Saga,$^{1}$\thanks{E-mail: shohei.saga@yukawa.kyoto-u.ac.jp}
Atsuhisa Ota,$^{2,3}$
Hiroyuki Tashiro$^{4}$
and Shuichiro Yokoyama$^{5,6}$
\\
$^{1}$Center for Gravitational Physics, Yukawa Institute for Theoretical Physics, Kyoto University, Kyoto 606-8502, Japan\\
$^{2}$Institute for Theoretical Physics and Center for Extreme Matter and Emergent Phenomena,\\
Utrecht University, Princetonplein 5, NL-3584 CC Utrecht, The Netherlands\\
$^{3}$Department of Applied Mathematics and Theoretical Physics, University of Cambridge, Cambridge, CB3 0WA, UK\\
$^{4}$Department of Physics and Astrophysics, Nagoya University, Nagoya, 464-8602, Japan\\
$^{5}$Kobayashi Maskawa Institute, Nagoya University, Aichi 464-8602, Japan\\
$^{6}$Kavli Institute for the Physics and Mathematics of the Universe (WPI),
Todai Institute for Advanced Study,\\
University of Tokyo, Kashiwa, Chiba 277-8568, Japan
}
\date{Accepted XXX. Received YYY; in original form ZZZ}

\pubyear{2019}

\begin{document}
\label{firstpage}
\pagerange{\pageref{firstpage}--\pageref{lastpage}}
\maketitle
\begin{abstract}
Spatially fluctuating primordial magnetic fields~(PMFs) inhomogeneously reheat the Universe when they dissipate deep inside the horizon before recombination.
Such an energy injection turns into an additional photon temperature perturbation.
We investigate secondary cosmic microwave background (CMB) temperature anisotropies originated from this mechanism, which we call {\it inhomogeneous magnetic reheating}.
We find that it can bring us information about non-linear coupling between PMFs and primordial curvature perturbations parametrized by $b_{\rm NL}$, which should be important for probing the generation mechanism of PMFs.
In fact, by using current CMB observations, we obtain an upper bound on the non-linear parameter as
$\log (b_{\rm NL} (B_{\lambda}/{\rm nG})^2) \lesssim {-36.5n_{B} - 94.0}$
with $B_{\lambda}$ and $n_{\rm B}$ being a magnetic field amplitude smoothed over $\lambda=1\; {\rm Mpc}$ scale and a spectral index of the PMF power spectrum, respectively.
Our constraints are far stronger than a previous forecast based on the future CMB spectral distortion anisotropy measurements because inhomogeneous magnetic reheating covers a much wider range of scales, i.e., $1\; {\rm Mpc}^{-1} \lesssim k\lesssim 10^{15}\; {\rm Mpc}^{-1}$.
\end{abstract}
\section{Introduction}

Intergalactic magnetic fields have recently become one of the most interesting and important topics in cosmology as several groups have reported \textit{lower} bounds on them, $B_0\gtrsim 10^{-15}$--$10^{-17}$\; G, by observing $\gamma$-ray emissions from blazars~(see e.g. \citealt{2011MNRAS.414.3566T,2012ApJ...747L..14V,Neronov:1900zz,Essey:2010nd,Chen:2014rsa}).
It is a puzzle that there exist such magnetic fields in intergalactic space, where there are few astrophysical objects.
One may answer the reason why magnetic fields exist there by introducing primordial magnetic fields (PMFs), which might be generated in the early Universe. 
Indeed, many authors have studied the mechanisms of generating PMFs from the viewpoint of high-energy physics in the early Universe (for reviews see e.g. \citealt{Widrow:2002ud,Grasso:2000wj,Durrer:2013pga,Subramanian:2015lua}), and the property of resultant PMFs highly depends on models of magnetogenesis.
Hence, it would be interesting that we may test high-energy physics through high-precision measurements of cosmological magnetic fields.

Among such observations, recent precise measurements of cosmic microwave background (CMB) anisotropies have provided rich information about PMFs~(see e.g. \citealt{Lewis:2004ef,2010PhRvD..81d3517S}).
PMFs imprint their signatures on the CMB anisotropies in various ways. On superhorizon scales, their anisotropic stress affects the geometry of the space-time and leads to the additional curvature perturbation and gravitational waves, which are called passive modes.
On sub-horizon scales, magnetohydrodynamical (MHD) effects would be significant.
Through these effects, the energy density and anisotropic stress of PMFs can set isocurvature scalar, vector, and tensor initial conditions, so-called, compensated modes~\citep{Lewis:2004ef,2012PhRvD..86d3510S,2010PhRvD..81d3517S}, and resultantly the MHD modes of tangled magnetic fields induce the CMB temperature and polarization anisotropies as discussed in~\citet{1998PhRvL..81.3575S}.
Moreover, PMFs also modify the thermal history of the intergalactic medium gas, which will also be imprinted in the CMB temperature anisotropies through the thermal Sunyaev--Zel'dovich effect~\citep{2017PhRvD..96l3525M}.

In this paper, we point out another effect on the CMB temperature anisotropies induced by PMFs.
In \citet{Saga:2017wwr}, the authors studied a late-time reheating mechanism through the diffusion of PMFs and estimated upper bounds on the PMF power spectrum on small scales.
Indeed, this idea was inspired by the works on acoustic reheating~\citep{Jeong:2014gna,Nakama:2014vla}.
Roughly speaking, acoustic reheating is heating due to acoustic damping.
Sound waves of a primordial baryon-photon plasma are dissipated due to shear viscosity on small scales.
This process leaves nothing in linear perturbation theory; however, the average component of the Universe would be slightly reheated at second order in the fluctuations. 
Thus, this effect is of second order in temperature perturbations, and hence, it is sensitive to the amplitude of primordial fluctuations on the corresponding damping scales.
\citet{Jeong:2014gna} and \citet{Nakama:2014vla} derived a constraint on the amplitude of short-wavelength density perturbations by comparing the radiation temperature at BBN and that at recombination.
Then, \citet{Saga:2017wwr} applied this method for energy injection due to the dissipation of PMFs.
\citet{2015JCAP...05..049N} extended the framework of acoustic reheating to spatially fluctuated one.
If primordial fluctuations are non-Gaussian, the cross-correlation between the primary CMB temperature anisotropies and the secondary one induced by acoustic reheating would be non-zero.
\citet{2015JCAP...05..049N}
showed that the observed CMB temperature anisotropies implicitly put upper bounds on extremely squeezed shapes of primordial non-Gaussianity because the secondary anisotropy should be subdominant.
On the analogy of acoustic reheating, we can extend global magnetic reheating proposed in \citet{Saga:2017wwr} to spatially fluctuated one, which we call {\it inhomogeneous magnetic reheating}.
The dissipation of spatially fluctuated PMFs leads to inhomogeneous energy injections into the CMB photons, which would be seen as additional anisotropies in the CMB.
Moreover, if PMFs are non-Gaussian, it might be a novel probe to explore such primordial non-Gaussianity in PMFs similar to anisotropic acoustic reheating.

The non-Gaussianity in PMFs is recently well studied in the context of inflationary magnetogenesis by several authors~\citep{2011PhRvD..84l3525C,2012PhRvD..85j3532M,2012PhRvD..85l3523B,2012PhRvD..86l3528J,2013JCAP...02..003J,2014JCAP...07..012N,2014JCAP...06..053F,2018JCAP...10..031C,2019JCAP...01..048C}.
In particular, if there is a coupling between an inflaton~$\phi$ and a gauge field strength $F_{\mu\nu}$ during inflation, generated PMFs would become non-Gaussian through the non-linear interaction.
For example, $f^{2}(\phi)F^{\mu\nu}F_{\mu\nu}$ coupling, where
$f^{2}(\phi)$ is a model-dependent function, is discussed in~\citet{2011PhRvD..84l3525C,2012PhRvD..85j3532M,2012PhRvD..85l3523B}, and such a primordial non-Gaussianity would be a key observable to distinguish models of magnetogenesis.
\citet{Shiraishi:2012xt} gave a constraint on this type of primordial non-Gaussianity through the two-point function of the CMB anisotropy originated from the passive mode.
\citet{Ganc:2014wia} has also investigated a potential role of future CMB spectral distortion measurements for constraining the non-Gaussianity of PMFs.
In this paper, we will show that inhomogeneous magnetic reheating can put a much stronger limit on it.

This paper is organized as follows.
In the next section, we introduce the dissipation mechanism of PMFs based on the MHD analysis and formulate inhomogeneous magnetic reheating as a straightforward extension of global magnetic reheating.
Then we discuss possible corrections to the CMB temperature anisotropies induced by PMFs in section~3.
In section~4, we compute the angular power spectrum and discuss the constraints on PMFs in section~5.
Finally, we will devote section~6 to summarize this paper.

\section{Inhomogeneous magnetic reheating}\label{sec: inhomogeneous mag reh}

\subsection{Dissipation of magnetic fields}
First of all, we review the dissipation mechanism of cosmological magnetic fields, following papers by \citet{Jedamzik:1996wp,Subramanian:1997gi}.
In the early universe, it was filled with highly conductive plasma.
Therefore, we may apply ideal MHD to analyze the evolution of the fluid
dynamics and magnetic fields.
It is also known that one can generalize MHD equations in flat space-time to those in the expanding Universe~\citep{1996PhRvD..54.1291B,Subramanian:1997gi}.
In the cosmological setup, we decompose PMFs into two scales: long-wavelength homogenous magnetic fields $\bm B_0$ and short-wavelength tangled ones $\bm B$.
We usually assume that we can take $|\bm B|\ll|\bm B_0|$ and linearize the Euler equation and ideal MHD equations to investigate subhorizon dynamics of the \textit{comoving} magnetic fields $\bm{b}(\tau,\bm{x}) = a^{2}(\tau)\bm{B}(\tau,\bm{x})$~\citep{Jedamzik:1996wp}.
It should be noticed that this assumption may not be easily justified for a given configuration of PMFs, but we avoid full non-linear analysis, which is beyond the scope of this paper.

In the linearized MHD, there are three types of modes: the fast and slow magnetosonic modes, and the Alfv\' en mode.
The fast and slow modes are literally sound waves in the magnetized plasma, and the Alfv\' en modes are incompressible waves.
Then, which MHD mode is excited by stochastic PMFs?
\citet{Subramanian:1997gi} and \citet{Jedamzik:1996wp} showed that excitation of MHD modes depends on the angle between $\bm{B}_0$ and propagation direction $\bm{k}$.
For example, if $\bm{B}_0 \parallel \bm{k}$, the magnetosonic modes become standard sound waves without magnetic pressure, and the tangled magnetic fields excite only the Alfv\'en modes, which satisfy $\bm{b}\perp \bm{B}_0$.
On the other hand, if $\bm{B}_0 \perp \bm{k}$, the Alfv\'en modes are not induced, and the MHD modes become the fast magnetosonic modes with $\bm{b}\parallel \bm{B}_0$.
In this case, the Fourier component\footnote{
The Fourier component of the comoving amplitude of PMFs is defined as
\begin{displaymath}
 \bm{b}(\tau,{\bm k}) \equiv \int {\rm d}^{3} x ~\bm{b}(\tau,{\bm x}) {\rm e}^{-{\rm i} {\bm k}\cdot {\bm x}}~.
\end{displaymath}
}
of the tangled magnetic field associated with the fast mode satisfies the following equation~\citep{Subramanian:1997gi}:
\begin{equation}
\frac{\partial^{2}{\bm{b}}(\tau,\bm{k})}{\partial\tau^{2}}
+ \frac{\eta }{a\bar{\rho}_{\rm r}} k^{2} \frac{\partial{\bm{b}}(\tau,\bm{k})}{\partial\tau} + k^{2}(c^{2}_{\rm s}+V_A^2){\bm{b}}(\tau,\bm{k}) = 0~, \label{eq: tangled}
\end{equation}
where $\tau$, $\bar{\rho}_{\rm r}$, $c_{\rm s}$, and $V_A$ are
the conformal time, the background energy density of radiation, the sound speed, and the Alfv\'en velocity, respectively.
Here $\eta = \frac{4}{15}\bar{\rho}_{\gamma}l_{\gamma}$ is shear viscosity emerges in photon-baryon fluids with $l_{\gamma} = (\sigma_{\rm T}n_{\rm e})^{-1}$ being the photon mean-free path.
Then, WKB solutions of Eq.~(\ref{eq: tangled}) become
\begin{equation}
\bm{b}(z,\bm{k}) = \bm{\tilde{b}}(\bm{k}) {\rm e}^{-(k/k_{\rm D}(z))^2} e^{\pm {\rm i} k \sqrt{c_{\rm s}^2+V_A^2} \tau} ~, \label{eq: split b}
\end{equation}
where $z$ is a cosmological redshift, $\bm{\tilde{b}}(\bm{k})$ denote the initial magnetic fields, and $k_{\rm D}$ expresses the damping scale of magnetic fields given by
\begin{equation}
k^{-2}_{\rm D} = \frac{2}{15}\int\frac{l_{\gamma}(t)}{a^{2}(t)}{\rm d}t ~,
\label{eq: damping scale}
\end{equation}
which is the same with the usual acoustic damping scale.
The time dependence of $k_{\rm D}$ can be read from Eq.~(\ref{eq: damping scale}) as $k^{-2}_{\rm D} \propto (1+z)^{-3}$.
Since the mean free path of photons becomes larger as the Universe evolves, the dissipation scale also increases.
Note that we only considered photons for shear viscosity, and this is valid for~$z\lesssim 10^{7}$.
In order to calculate the damping scale in $z\gtrsim 10^{7}$, we need to consider neutrino free streaming.
Moreover, anisotropic shear of relativistic weak bosons would be significant for~$z\gtrsim 10^{15}$.
Here we account for these effects, following~\citet{Jeong:2014gna}.

The above two cases, i.e., $\bm{B}_0 \parallel \bm{k}$ and $\bm{B}_0 \perp \bm{k}$ are specific choices of propagation directions.
These MHD modes are mixed in the case of the intermediate configurations, and it is not straightforward to analyse the dispersion relations.
Thus, the excitation of MHD modes depends on realizations of $\bm{b}$ on top of $\bm{B}_0$ and their propagation directions.
Since the initial conditions of MHD modes would be randomly given for stochastic PMFs, we may, therefore, assume the equipartition of PMFs for the initial conditions of each MHD mode.

As we have mentioned, the diffusion scale of the fast mode is determined by the damping scale of sound waves.
This implies that the fast modes always travel with sound speed if Alfv\'en velocity is negligible, and the dissipative feature does not depend on the size of background magnetic fields.
In contrast, the dissipations of the slow and Alfv\'en modes which propagate with Alfv\'en velocity, are more non-trivial.
If the $B_{0}$ is tiny, the slow and Alfv\'en modes propagate so slowly that they are overdamped.
In this case, a driving force of oscillations is balanced with viscosity, and then dissipation of these modes becomes inefficient~\citep{Subramanian:1997gi,Jedamzik:1996wp}.
Thus, the properties of slow modes and Alfv\'en modes highly depend on the significance of background magnetic fields. 
Hence we ignore the slow and Alfv\'en modes, and we assume that the background magnetic fields are weak enough to have $V_A\ll c_s$ for simplicity.
Dropping these two modes forces us to underestimate the effects
of magnetic reheating.
However, it would only change an $O(1)$ factor to our results.

\subsection{Energy injection into CMB photons}
As we have discussed in the previous subsection, a randomly given fast magnetosonic mode is damping due to shear viscosity of relativistic particles on small scales, regardless of the size of background magnetic fields.
Then the energy of such a magnetic field in a comoving volume is decreasing.
Suppose the Universe is dominated by radiation, the released energy would be redistributed to photons by the Compton scattering.
Below, we compute temperature variations due to inhomogeneous magnetic reheating, solving the conservation laws in the total system of radiations and PMFs.

First of all, we write the comoving energy density of PMFs
\begin{align}
    \rho_{B}(z, \bm{x}) = \frac{1}{8\pi}|\bm{b}(z,\bm{x})|^2. \label{eq: injection mag}
\end{align}
The damping effect is taken into account in the evolution of $\bm{b}(z,{\bm x})$, as seen in Eq.~(\ref{eq: split b}).
According to Eqs.~\eqref{eq: split b} and \eqref{eq: injection mag}, we can write the energy density in Fourier space as
\begin{align}
\rho_{B}(z, \bm{k}) &=
\frac{1}{8\pi}\int\frac{{\rm d}^{3}k_{1}{\rm d}^{3}k_{2}}{(2\pi)^{3}}
\delta^{3}_{\rm D}(\bm{k - \bm{k}_{1} -\bm{k}_{2}})
\notag \\
& \times 
\tilde{b}_{i}(\bm{k}_{1})\tilde{b}_{i}(\bm{k}_{2})
{\rm e}^{- \left(k^{2}_{1}+k^{2}_{2} \right)/k^{2}_{\rm D}(z)}
~. \label{eq:inhomo-dissi-rate}
\end{align}
As can be seen in Eqs. \eqref{eq: injection mag} and \eqref{eq:inhomo-dissi-rate}, the energy density of PMFs would spatially fluctuate and hence
the magnetic reheating ratio, i.e., energy injection into the CMB photons should also fluctuate.

For $z \gtrsim z_{\mu} =  2.0\times 10^{6}$, the Universe is in chemical equilibrium by the Compton scattering, bremsstrahlung, the double Compton scattering, etc. (e.g., \citealt{1982A&A...107...39D,1970Ap&SS...7....3S,1991A&A...246...49B,1975SvA....18..413I}), so that the photon distribution function becomes a Planck distribution function of the form $\left[\exp(p/T_0(1+\Delta_B))-1\right]^{-1}$, with a comoving temperature $T_0$ and a dimensionless magnetic reheating temperature rise $\Delta_B$.
Note that we dropped the temperature perturbations originated from primordial density perturbations, but it would be straightforward to add such a contribution.
During this period, variations of photon temperature are given by the conservation law of energy.
The photon energy density is given by
\begin{align}
    \rho_\gamma &\approx \rho_{\gamma,0} \left(1+4\Delta_B\right),
\end{align}
where $\rho_{\gamma,0}\propto T_0^4$ is the comoving energy density of photons without magnetic reheating.
Then, the conservation law
\footnote{
In the early Universe, the effective relativistic degrees of freedom are larger than those of photons. Therefore, it should be noticed that photons' share of energy injection depends on when magnetic reheating happens.
However, we ignore this effect for simplicity, because the number of relativistic species in the very early Universe is still an open question.}
of the total energy in a local diffusion patch around $\bm{x}$ is
\begin{align}
    {\rm d}(\rho_\gamma +\rho_B) = 0.\label{planck:mh}
\end{align}
Note that the number of photons are not conserved when the system is in chemical equilibrium.
Solving Eq.~\eqref{planck:mh}, we immediately find
\begin{equation}
4{\rm d} \Delta_{B} \approx  
 - \frac{{\rm d} \rho_{B}}{\rho_{\gamma,0}}  ~. \label{planck:eq: temp mag}
\end{equation}
Thus, an additional temperature perturbation would be given by 1/4 of local energy injection at leading order.

On the other hand, bremsstrahlung and the double Compton scattering become less efficient below $z_{\mu}$.
Then, photon number cannot be changed, but photons are, still, in kinetic equilibrium due to the efficient Compton scattering.
Hence energy injection not only raises the temperature but also produces the chemical potential called $\mu$-distortion to respect the conservation laws of both energy and number~\citep{Chluba:2012gq}.
In contrast to a Planck distribution function, it should be noticed that the temperature of a Bose--Einstein distribution function is not the fourth root of energy.
Given a Bose--Einstein distribution $\left[\exp(p/T_0(1+\Delta_B)+\mu)-1\right]^{-1}$, with a dimensionless chemical potential $\mu$, we get the following comoving energy and number density at leading order in $\mu$ and $\Delta_B$:
\begin{align}
    \rho_\gamma &\approx \rho_{\gamma,0} \left(1+4\Delta_B -\frac{90\zeta(3)}{\pi^4}  \mu \right) ~, \\
    n_\gamma &\approx n_{\gamma,0} \left(1+3\Delta_B -\frac{\pi^2}{6\zeta(3)}  \mu \right) ~,
\end{align}
where $n_{\gamma,0}\propto T_0^3$ is number density of photons without magnetic reheating, and $\zeta(n)$ are the Riemann zeta functions.
Then, the conservation laws of the comoving energy and number density in a local diffusion patch around $\bm{x}$ would be given as
\begin{align}
    {\rm d}(\rho_\gamma+\rho_B)=0,~~ {\rm d}n_\gamma =0,
\end{align}
which can be recast into
\begin{align}
    4{\rm d}\Delta_B  -\frac{90\zeta(3)}{\pi^4}{\rm d}\mu &= -\frac{{\rm d}\rho_B}{\rho_{\gamma,0}} ~,\\
    3{\rm d}\Delta_B  -\frac{\pi^2}{6\zeta(3)}{\rm d}\mu &= 0.
\end{align}
Eliminating $\mu$, one finds
\begin{align}
    4{\rm d}\Delta_B \approx -\alpha \frac{{\rm d}\rho_B}{\rho_{\gamma,0}}~,
\end{align}
where $\alpha = \left( 1 - 405\left( \zeta(3) \right)^{-1}/\pi^{6}\right)^{-1} \approx 2.56$.

After $z_{\rm f}  = 4\times 10^{4}$ the Universe is out of kinetic equilibrium; therefore, energy injection is not trivially transferred to photons, and hence we ignore this period for simplicity \citep{1970Ap&SS...7....3S}.
To summarize, we can describe the local heating rate due to the PMF dissipation
as
\begin{equation}
\frac{{\rm d} \Delta_{B}(z, \bm{x})}{{\rm d} z} = \frac{1}{4} 
\mathcal{M}(z)
\left[ - \frac{1}{\rho_{\gamma,0}} \frac{{\rm d} \rho_{B}}{{\rm d} z}(z,\bm{x}) \right] ~, \label{eq: temp mag}
\end{equation}
where we introduced $\mathcal{M}(z)$ as
\begin{equation}
\mathcal{M}(z) =
\begin{cases}
1 ~~~ ( z_{\mu} \leq z \leq z_{\rm i}) \\
\alpha ~~~ (z_{\rm f} \leq z \leq z_{\mu}) ~.
\end{cases} 
\end{equation}
Here $z_{\rm i}$ is the redshift of generation of PMFs,
which depends on the magnetogenesis scenario.
Throughout this paper, we set $z_{\rm i} = 10^{15}$~(corresponding to the electroweak phase transition era).
The PMF dissipation does not depend on CMB photon's direction $\hat{\bm{n}}$.
Therefore, inhomogeneous magnetic reheating acts as a monopole source term in the Boltzmann hierarchical equations for CMB photons.

\section{Temperature anisotropy induced by PMFs}
\label{Temperaturepmfs}

In this section, we discuss corrections to the CMB temperature anisotropy in the presence of PMFs.
Let us decompose the temperature harmonic coefficients into the following three parts:
\begin{equation}
a_{\ell,m} = a^{\Theta}_{\ell,m}+a^{\Theta_{B}}_{\ell,m}+a^{\Delta_{B}}_{\ell,m}~.
\end{equation}
The first term is the standard linear temperature anisotropies given as
\begin{equation}
a^{\Theta}_{\ell,m} = 4\pi {\rm i}^{\ell} \int\frac{{\rm d}^{3} k}{(2\pi)^{3}} Y^{*}_{\ell ,m}(\hat{\bm{k}}) \mathcal{T}_{\ell}(k) \left( \frac{3}{5}\zeta(\bm{k})\right)
~, \label{eq: alm adiabatic}
\end{equation}
where $\zeta$ is the primordial curvature perturbation on the uniform density slice and $\mathcal T_\ell$ are the transfer functions of the temperature perturbations.
On the other hand, $a^{\Theta_{B}}_{\ell,m}$ are the additional anisotropies due to the passive mode and $a^{\Delta_{B}}_{\ell,m}$ are the new corrections of inhomogeneous magnetic reheating which we will define later.
We observe the total angular power spectrum
\begin{equation}
    C_{\ell} = \frac{1}{2\ell + 1}\sum_{m} \Braket{a_{\ell,m} \left( a_{\ell,m}\right)^{*}} ~,
\end{equation}
which contains the auto- and cross-correlations 
\begin{equation}
C^{XY}_{\ell}
=
\frac{1}{2\ell + 1}\sum_{m} \Braket{a^{X}_{\ell,m} \left( a^{Y}_{\ell,m}\right)^{*}} ~,
\end{equation}
where $X/Y = \Theta, \Delta_{B}$, or $\Theta_{B}$.
The observed temperature anisotropies are consistent with $C^{\Theta\Theta}_{\ell}$.
Therefore, the corrections due to PMFs should be subdominant so that we can put upper bounds on both the passive modes and magnetic reheating.
Let us see the explicit expressions of the harmonic coefficients in the following subsections.
\footnote{On scales smaller than the horizon scale at the recombination
epoch, PMFs might induce isocurvature-like perturbations and
generate the additional CMB anisotropies which are called as compensated
magnetic modes~\citep{Lewis:2004ef,2012PhRvD..86d3510S,2010PhRvD..81d3517S}. However, these anisotropies depend on the
initial conditions, in particular, the relation between matter and
the model of magnetogenesis. Therefore, we neglect these types of CMB anisotropies. }

\subsection{Passive mode}

PMFs intrinsically have anisotropic stress, which can induce an additional curvature perturbation $\zeta_B$ on superhorizon scales.
It would subsequently imprint on the CMB temperature anisotropy.
Note that neutrino anisotropic stress cancels that of the magnetic fields; the therefore generation of $\zeta_B$ stops after neutrino decoupling.
The multipole coefficient of the passive mode is then given as~\citep{2010PhRvD..81d3517S}
\begin{equation}
a^{\Theta_{B}}_{\ell,m} = 4\pi {\rm i}^{\ell}\int\frac{{\rm d}^{3}k}{(2\pi)^{3}}
Y^{*}_{\ell ,m}(\hat{\bm{k}}) \mathcal{T}_{\ell}(k) \frac{3}{5} \zeta_{B}(\bm{k})
~.\label{eq: alm passive mode}
\end{equation}
Here $\zeta_{B}$ is the curvature perturbation generated by PMFs,
\begin{equation}
\zeta_{B}(\bm{k}) = \frac{1}{3}\xi R_{\gamma}\Pi_{B}(\bm{k}) ~,\label{eq:passive mode}
\end{equation}
where we define $\xi = \ln{\left( {\eta_{\nu}}/{\eta_{B}}\right)} +
{5}/{(8R_{\nu})} - 1$, $R_{\gamma} = \rho_{\gamma}/(\rho_{\gamma} +
\rho_{\nu})$, $R_{\nu} = 1 - R_{\gamma}$, $\eta_{\nu}$ is the conformal
time at the neutrino decoupling epoch, $\eta_{B}$ is the conformal time when PMFs are generated, and $\Pi_{B}$ is the scalar part of the PMF anisotropic stress, which is obtained as
\begin{align}
\Pi_{B}(\bm{k}) &= \frac{9}{8\pi \rho_{\gamma,0}} \int\frac{{\rm d}^{3}k_{1}}{(2\pi)^{3}}\int\frac{{\rm d}^{3}k_{2}}{(2\pi)^{3}}
(2\pi)^{3}\delta^{3}_{\rm D}(\bm{k}-\bm{k}_{1}-\bm{k}_{2})
\notag \\
&\times 
\left( \hat{k}_{i}\hat{k}_{j} - \frac{1}{3}\delta_{ij}\right)C_{\Theta_B}(k_{1},k_{2})\tilde{b}_{i}(\bm{k}_{1})\tilde{b}_{j}(\bm{k}_{2})
~.\label{eq:passive mode:P}
\end{align}
In the above equation, we introduced the damping scale of PMFs, following~\citet{2010PhRvD..81d3517S,2016A&A...594A..19P} as
\begin{equation}
C_{\Theta_B}(k_{1},k_{2}) = {\rm e}^{-\left( k^{2}_{1}+k^{2}_{2}\right)/k^{2}_{\rm D}(z_{\nu})} ~, \label{eq: C(k,k) theta B}
\end{equation}
where $z_{\nu}$ is the redshift at the neutrino decoupling epoch.
Note that \citet{2010PhRvD..81d3517S} dropped time dependence of $\Pi_B$ when they derived Eq.~(\ref{eq:passive mode}) by integrating the Einstein equation, and they evaluated $\Pi_B$ at neutrino decoupling.
This simplification makes us underestimate the contribution of the passive modes, but we justify this prescription because the relevant scales are covered by magnetic reheating as we will show in Fig.~\ref{fig: C k}.

\subsection{Magnetic reheating}

For magnetic reheating, integrating Eq.~(\ref{eq: temp mag}) with respect to the redshift, we get 
\begin{align}
\Delta_{B}(\bm{k}) &= 
\frac{1}{32\pi \rho_{\gamma, 0}}
\int\frac{{\rm d}^{3}k_{1}}{(2\pi)^{3}}
\int\frac{{\rm d}^{3}k_{2}}{(2\pi)^{3}}
(2\pi)^{3}\delta^{3}_{\rm D}(\bm{k}-\bm{k}_{1}-\bm{k}_{2})
\notag \\
& \times
C_{\Delta_{B}}(k_{1},k_{2})
\tilde{b}_{i}(\bm{k}_{1}) \tilde{b}_{i}(\bm{k}_{2})
~,  \label{eq: T Fourier space}
\end{align}
where $C_{\Delta_{B}}(k_{1},k_{2})$ is a kernel function related to the dissipation efficiency of PMFs as
\begin{align}
C_{\Delta_{B}}(k_{1}, k_{2}) & =
\exp\left[ -\frac{k^{2}_{1} + k^{2}_{2}}{k^{2}_{\rm D}(z_{\rm i})}\right]
-  \exp\left[ -\frac{k^{2}_{1} + k^{2}_{2}}{k^{2}_{\rm D}(z_{\mu})}\right]
\notag \\
& +
\alpha\left( \exp\left[ -\frac{k^{2}_{1} + k^{2}_{2}}{k^{2}_{\rm D}(z_{\mu})}\right]
-  \exp\left[ -\frac{k^{2}_{1} + k^{2}_{2}}{k^{2}_{\rm D}(z_{\rm f})}\right]  \right)
~. \label{eq: def C(k,k)}
\end{align}
Then we find that the resultant CMB angular power spectrum induced by magnetic reheating can be expressed by the following approximate formula for the multipole coefficient:
\begin{equation}
a^{\Delta_{B}}_{\ell,m} \approx  4\pi {\rm i}^{\ell} \int\frac{{\rm d}^{3}k}{(2\pi)^{3}} Y^{*}_{\ell ,m}(\hat{\bm{k}}) \mathcal{T}_{\ell}(k) \Delta_{B}(\bm{k}) ~. \label{eq: alm Fourier}
\end{equation}

\section{Corrections to the CMB angular power spectrum}

We are now in a position to calculate the PMF corrections to the CMB temperature
angular power spectrum~$C_{\ell}$.
In this work, we compute~$C_{\ell}$ by modifying public Boltzmann codes,~e.g.,~{\tt CLASS}~\citep{2011arXiv1104.2932L}.
According to Eqs.~\eqref{eq: alm adiabatic},~\eqref{eq: alm passive mode}
and \eqref{eq: alm Fourier},
it should be noticed that $C^{XY}_{\ell}$ can contain two types of initial correlation functions: $\langle B^4\rangle$ and $\langle B^2\zeta\rangle$.
The four-point function of $B$ can be reduced to products of $\langle B^2\rangle $ at the leading order.
On the other hand, the latter correlation is non-zero in the case where the coupling between PMFs and the primordial curvature perturbations exists.
We will see that the scale dependence is entirely different for these two initial conditions in the following subsections.

\subsection{Gaussian disconnected part: $\Braket{B^{4}}$}
\label{sec:Gaussian}

When initial PMFs are statistically homogeneous and isotropic Gaussian random fields,
the stochastic property of PMFs is completely characterized by the power spectrum
\begin{equation}
\braket{\tilde{b}_{i}(\bm{k})\tilde{b}^{*}_{j}(\bm{k'})} = (2\pi)^{3}\delta^{3}_{\rm D}(\bm{k} - \bm{k'}) \frac{1}{2} \left( \delta_{ij} - \hat{k}_{i}\hat{k}_{j}\right) P_{B}(k) ~.\label{mg:power}
\end{equation}
Then, the four-point function of $B$ is reduced to the product of $P_{B}$.
Taking the ensemble averages of Eqs.~(\ref{eq: alm Fourier}) and (\ref{eq: alm passive mode}),
we obtain the following angular power spectrum of auto and cross-correlation
\begin{align}
C^{\Delta_{B}\Delta_{B}}_{\ell} & =
\frac{1}{128 \rho^{2}_{\gamma, 0}} \frac{1}{(2\pi)^{5}}
\int{ {\rm d}k}\, k^{2}
\mathcal{T}^{2}_{\ell} (k)
\int{{\rm d} k_{1}}\, k^{2}_{1}
\int^{1}_{-1}{\rm d}\mu\;
\notag \\
&\times 
\mathcal{F}(k,k_{1},\mu)
P_{B}(k_{1})P_{B}(|{\bm k}-{\bm k}_1|)
\left( C_{\Delta_{B}}(k_{1}, |{\bm k}-{\bm k}_1|) \right)^{2}
~, \label{eq: cl auto} \\
C^{\Theta_B \Delta_{B}}_{\ell} & =
\frac{9R_{\gamma}\xi}{160 \rho^{2}_{\gamma, 0}} \frac{1}{(2\pi)^{5}}
\int{ {\rm d}k}\, k^{2}
\mathcal{T}^{2}_{\ell} (k)
\int{{\rm d} k_{1}}\, k^{2}_{1}
\int^{1}_{-1}{\rm d}\mu\;
\mathcal{G}(k,k_{1},\mu)
\notag \\
& \times 
P_{B}(k_{1})P_{B}(|{\bm k}-{\bm k}_1|)
C_{\Theta_B}(k_{1}, |{\bm k}-{\bm k}_1|)
C_{\Delta_{B}}(k_{1}, |{\bm k}-{\bm k}_1|)
~, \label{eq: cl cross}
\\
C^{\Theta_{B}\Theta_{B}}_{\ell}
& =
\frac{81 R^{2}_{\gamma}\xi^{2}}{200\rho^{2}_{\gamma,0}}
\frac{1}{(2\pi)^{5}}
\int{{\rm d}k}\; k^{2}
\mathcal{T}^{2}_{\ell}(k) 
\int{{\rm d}k_{1}}\; k^{2}_{1}
\int^{1}_{-1}{\rm d}\mu\; 
\notag \\
&\times 
\mathcal{I}(k,k_{1},\mu)
P_{B}(k_{1}) P_{B}(|{\bm k}-{\bm k}_1|)
\left( C_{\Theta_B}(k_{1}, |{\bm k}-{\bm k}_1|) \right)^{2}
~,
\label{eq: cl passive}
\end{align}
where $\mu = \hat{\bm{k}}\cdot \hat{\bm{k}}_{1}$ and the configuration factor, $\mathcal{F}(k,k_{1},\mu)$, $\mathcal{G}(k,k_{1},\mu)$, and $\mathcal{I}(k,k_{1},\mu)$ are given by
\begin{align}
\mathcal{F}(k,k_{1},\mu)
&=
\frac{(1+\mu^{2})k^{2} - 4k k_{1} \mu+ 2k^{2}_{1}}{k^{2} + k^{2}_{1} - 2k k_{1} \mu} ~, \label{eq: calF} \\
\mathcal{G}(k,k_{1},\mu)
&= \frac{1}{3}\frac{k^{2}_{1}(1-3\mu^{2}) - k^{2}(1+\mu^{2}) + k k_{1}\mu (1+3\mu^{2})}{k^{2} + k^{2}_{1} - 2k k_{1} \mu} ~, \\
\mathcal{I}(k,k_{1},\mu)
& =
\frac{1}{9}\frac{k^{2}(1+\mu^{2}) + k k_{1}(2\mu - 6\mu^{3}) + k^{2}_{1}(5-12\mu^{2}+9\mu^{4})}{k^{2} + k^{2}_{1} - 2k k_{1} \mu} ~.
\end{align}

Let us substitute the following delta-function type power spectrum of PMFs
\begin{equation}
P_{B}(\ln{k}) = \frac{2\pi^{2}}{k^{3}}\mathcal{B}^{2} \delta_{\rm D}\left( \ln\left( k/k_{\rm p}\right) \right) ~. \label{eq: delta power}
\end{equation}
Using this power spectrum, we would find the angular scale which is sensitive to a given $k_{\rm p}$ mode.
We can proceed $\mu$- and $k_{1}$-integrals appeared in Eqs.~(\ref{eq: cl auto}), (\ref{eq: cl cross}), and (\ref{eq: cl passive}) as
\begin{align}
C^{\Delta_{B}\Delta_{B}}_{\ell}
& \approx 
\frac{\mathcal{B}^{4}}{8\pi\rho^{2}_{\gamma,0}}
\frac{1}{64}
\left( C_{\Delta_{B}}(k_{\rm p}, k_{\rm p}) \right)^{2}
\int\frac{{\rm d}k}{k}\; 
\mathcal{T}^{2}_{\ell} (k)
\frac{k^{2}}{k^{2}_{\rm p}} 
\label{eq: Cl delta BB approx}
~,\\
C^{\Theta_{B}\Delta_{B}}_{\ell}
& \approx 
\frac{\mathcal{B}^{4}}{8\pi\rho^{2}_{\gamma,0}}
\frac{3R_{\gamma}\xi}{160}
C_{\Theta_B}(k_{\rm p}, k_{\rm p})
C_{\Delta_{B}}(k_{\rm p}, k_{\rm p})
\int\frac{{\rm d}k}{k}\;
\mathcal{T}^{2}_{\ell} (k)
\frac{k^{2}}{k^{2}_{\rm p}} 
~,\\
C^{\Theta_{B}\Theta_{B}}_{\ell}
&\approx
\frac{\mathcal{B}^{4}}{8\pi\rho^{2}_{\gamma,0}}
\frac{9 R^{2}_{\gamma}\xi^{2}}{40}
\left( C_{\Theta_B}(k_{\rm p}, k_{\rm p}) \right)^{2}
\int\frac{{\rm d}k}{k}\;
\mathcal{T}^{2}_{\ell}(k)
\frac{k^{2}}{k^{2}_{\rm p}}
~,
\end{align}
where we use the fact that the transfer function can be approximated as
$\mathcal{T}_{\ell}(k) \propto j_{\ell}(k\eta_{0})$ with the present conformal time $\eta_{0}$.
The integration over $k$
can pick up the contribution only $k_{\rm p}\gg k \sim \ell /\eta_{0}$
with the configuration factor $\mathcal{F}\approx 2$, $\mathcal{G}\approx
1/3$, and $\mathcal{I}\approx 5/9$.

\begin{figure}
\includegraphics[width=\columnwidth]{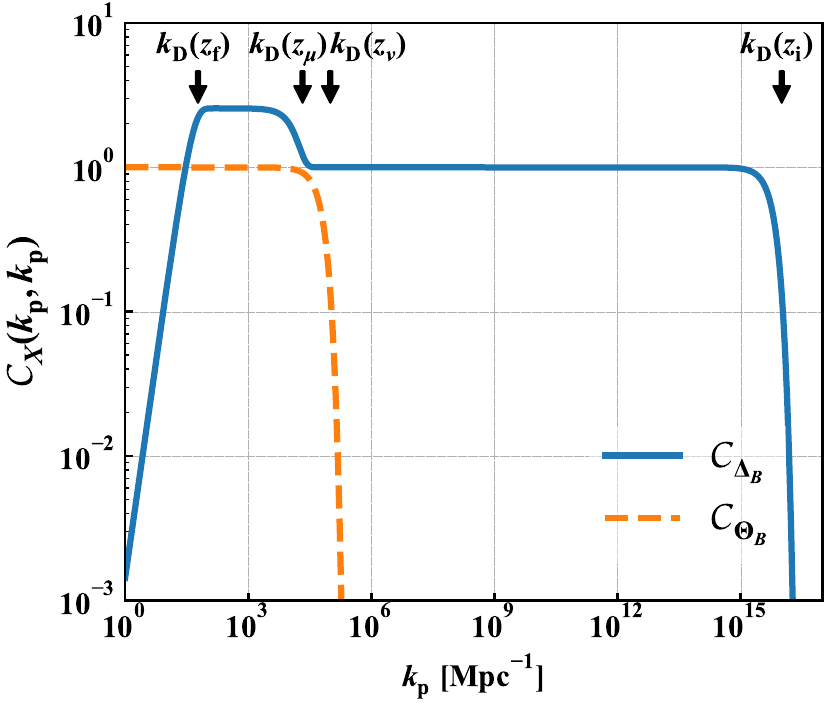}
\caption{The kernel functions $C_{\Delta_{B}}(k_{\rm p},k_{\rm p})$ (blue, solid) and $C_{\Theta_B}(k_{\rm p},k_{\rm p})$ (orange, dashed) related to inhomogeneous magnetic reheating and passive mode, respectively, as a function of $k_{\rm p}$, defined in Eqs.~(\ref{eq: def C(k,k)}) and (\ref{eq: C(k,k) theta B}), respectively.
The arrows show the diffusion scales of each redshift, i.e., 
$k_{\rm D}(z_{\rm f}) \approx 6\times 10^{1} \; {\rm Mpc}^{-1}$,
$k_{\rm D}(z_{\mu}) \approx 2 \times 10^{4} \; {\rm Mpc}^{-1}$,
$k_{\rm D}(z_{\rm \nu}) \approx 1.0\times 10^{5} \; {\rm Mpc}^{-1}$,
and $k_{\rm D}(z_{\rm i}) \approx 10^{16} \; {\rm Mpc}^{-1}$ based on \citet{Jeong:2014gna}.
}
\label{fig: C k}
\end{figure}

We find that $k_{\rm p}$ dependence is summarised in $C_{\Delta_{B}}(k_{\rm p},k_{\rm p})$ and $C_{\Theta_B}(k_{\rm p},k_{\rm p})$, which are shown in Fig.~\ref{fig: C k}.
We can see that the shape of $C_{\Delta_{B}}(k_{\rm p},k_{\rm p})$
appears like a rectangular function (blue, solid),
which can explore much smaller scales than the passive mode
shown in the orange dashed line.
Fig.~\ref{fig: C k} shows that
both inhomogeneous magnetic reheating and
the passive mode can probe to the PMFs
on the scales, $1~{\rm Mpc^{-1}} < k_{\rm p}< 10^4~{\rm Mpc^{-1}}$.
On the other hand,
on smaller scales, i.e., $k_{\rm p} \gg O(10^4)\; {\rm Mpc}^{-1}$, contrary to the passive mode, inhomogeneous magnetic reheating is possibly available to constrain the amplitude of PMFs.

Note that, from the integrand in Eq.~(\ref{eq: Cl delta BB approx}), we find that the shape of the angular power spectrum $C^{\Delta_{B}\Delta_{B}}_{\ell}$ corresponds to the one induced by the adiabatic initial condition with a scalar spectral index $n_{\rm s} = 3$.

\subsection{Non-Gaussian connected part: $\Braket{B^{2}\zeta}$}

Several magnetogenesis models, for example,
$f^{2}(\phi)F^{\mu\nu}F_{\mu\nu}$ interaction, produce non-zero
cross-correlation between the energy density of PMFs and the curvature perturbations (e.g.,~\citealt{2011PhRvD..84l3525C,2012PhRvD..85j3532M,Shiraishi:2012xt}).
In such a case, we can parametrize the cross-correlation in the squeezed limit ($k_1 \sim k_2 \gg k_3$) following in \citet{2012PhRvD..86l3528J,Ganc:2014wia} as
\begin{align}
&\Braket{\tilde{b}_{i}(\bm{k}_{1})\tilde{b}_{j}(\bm{k}_{2})\zeta^{*}(\bm{k}_{3})}_{k_{3}\ll k_{1}\sim k_{2}} \notag \\
&=
(2\pi)^{3}\delta^{3}_{\rm D}(\bm{k}_{1} + \bm{k}_{2} - \bm{k}_{3})
 b_{\rm NL} \frac{\delta_{ij} - \hat{k}_{1i}\hat{k}_{1j}}{2}
P_{B}(k_{1})P_{\zeta}(k_{3})
~,
\label{eq: def bNL}
\end{align}
where $P_{\zeta}(k)$ is a power spectrum of the primordial curvature perturbations, which is simply parametrized as
\begin{equation}
\frac{k^{3}}{2\pi^{2}}P_{\zeta}(k) = \mathcal{A}_{\zeta}\left( \frac{k}{k_{0}}\right)^{n_{\rm s}-1}~.
\label{eq: Pzeta}
\end{equation}
In such a non-Gaussian case, superhorizon curvature perturbations couple to subhorizon magnetic fields.
Hence, inhomogeneous magnetic reheating on tiny scales leads to additional superhorizon temperature perturbations.
Inhomogeneous magnetic reheating is more sensitive to smaller scales than passive modes, as seen in Fig.~\ref{fig: C k}; therefore $C^{\Theta \Delta_B}_\ell$ would be more significant than $C^{\Theta\Theta_B}_\ell$.
Thus inhomogeneous magnetic reheating can become a novel probe of the correlation function~(\ref{eq: def bNL}).

The cross-correlations are given by using Eqs.~(\ref{eq: alm Fourier}), (\ref{eq: alm passive mode}), and (\ref{eq: alm adiabatic}) as
\begin{align}
C^{\Theta_{B}\Theta}_{\ell}
&= 
- b_{\rm NL}
\frac{27\pi \xi R_{\gamma}}{25 \rho_{\gamma,0}}
\frac{1}{(2\pi)^{5}}
\int{\rm d}k\; k^{2}\mathcal{T}^{2}_{\ell}(k)
\notag \\
&\times 
\int{\rm d}k_{1}\; k^{2}_{1}
\tilde{C}_{\Theta_B}(k,k_{1})
P_{B}(k_{1})P_{\zeta}(k)
~, \label{eq: bNL Bzeta zeta}
\end{align}
for the cross-correlation between the passive mode and adiabatic curvature perturbation, and
\begin{align}
C^{\Delta_{B} \Theta}_{\ell}
&= 
b_{\rm NL}
\frac{3\pi}{10 \rho_{\gamma,0}}
\frac{1}{(2\pi)^{5}}
\int{\rm d}k\; k^{2}\mathcal{T}^{2}_{\ell}(k)
\notag \\
&\times 
\int{\rm d}k_{1}\; k^{2}_{1}
\tilde{C}_{\Delta_{B}}(k,k_{1})
P_{B}(k_{1})P_{\zeta}(k) ~, \label{eq: bNL B zeta}
\end{align}
for the cross-correlation between inhomogeneous magnetic reheating and the adiabatic curvature perturbation.
In the above equations, we define
\begin{align}
\tilde{C}_{\Theta_B}(k,k_{1})
&= \int^{1}_{-1}{\rm d}\mu\; C_{\Theta_B}(k_{1}, |\bm{k} - \bm{k}_{1}|) \left( \mu^{2} - \frac{1}{3} \right)
\notag \\
&=
- \frac{k^{2}_{\rm D}(z_\nu)}{12k^{3}k^{3}_{1}}
\Biggl[
4 \left( {\rm e}^{- \frac{k^{2}_{+}}{k^{2}_{\rm D}(z_{\nu})}} - {\rm e}^{- \frac{k^{2}_{-}}{k^{2}_{\rm D}(z_{\nu})}}\right)
k^{2}k^{2}_{1}
\notag \\
& + 6 
 \left( {\rm e}^{- \frac{k^{2}_{+}}{k^{2}_{\rm D}(z_{\nu})}} + {\rm e}^{- \frac{k^{2}_{-}}{k^{2}_{\rm D}(z_{\nu})}}\right)
 k k_{1} k^{2}_{\rm D}(z_\nu)
\notag \\
&+ 3 
 \left( {\rm e}^{- \frac{k^{2}_{+}}{k^{2}_{\rm D}(z_{\nu})}} - {\rm e}^{- \frac{k^{2}_{-}}{k^{2}_{\rm D}(z_{\nu})}}\right)
 k^{4}_{\rm D}(z_\nu )
\Biggr]
~, \\
\tilde{C}_{\Delta_{B}}(k,k_{1})
&= \int^{1}_{-1}{\rm d}\mu\; C_{\Delta_{B}}(k_{1}, |\bm{k} - \bm{k}_{1}|) \notag \\
&=
\frac{1}{6k k_{1}}
\Biggl[
3\alpha
\left( {\rm e}^{- \frac{k^{2}_{+}}{k^{2}_{\rm D}(z_{\rm f})}}
- {\rm e}^{- \frac{k^{2}_{-}}{k^{2}_{\rm D}(z_{\rm f})}}\right) k^{2}_{\rm D}(z_{\rm f}) \notag \\
& - 3
\left( {\rm e}^{- \frac{k^{2}_{+}}{k^{2}_{\rm D}(z_{\rm i})}} 
- {\rm e}^{- \frac{k^{2}_{-}}{k^{2}_{\rm D}(z_{\rm i})}}\right) k^{2}_{\rm D}(z_{\rm i}) \notag \\
& + 3(1-\alpha)\left( {\rm e}^{- \frac{k^{2}_{+}}{k^{2}_{\rm D}(z_{\mu})}}
- {\rm e}^{- \frac{k^{2}_{-}}{k^{2}_{\rm D}(z_{\mu})}} \right) k^{2}_{\rm D}(z_{\mu})
\Biggr] ~,
\end{align}
where we define $k^{2}_{\pm} = k^{2} \pm 2k k_{1} + 2k^{2}_{1}$.

\begin{figure}
\begin{center}
\includegraphics[width=\columnwidth]{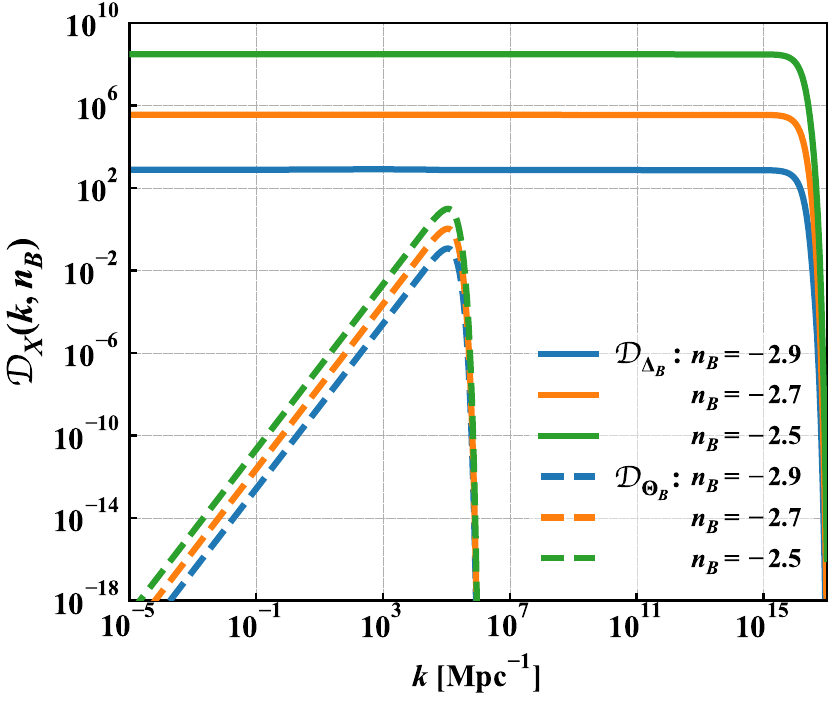}
\includegraphics[width=\columnwidth]{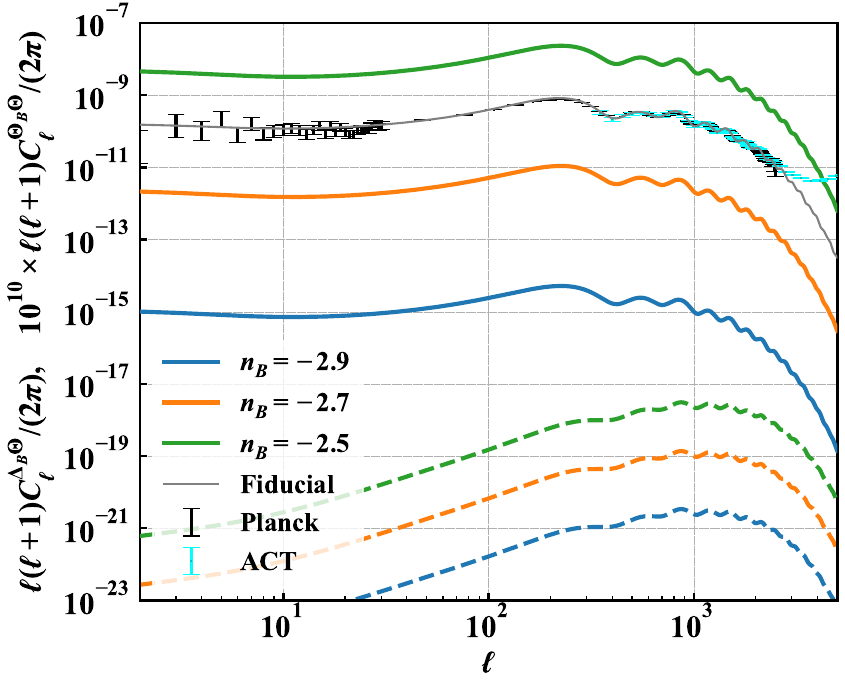}
\end{center}
\caption{({\it Top}) The functions $\mathcal{D}_{\Delta_{B}}(k,n_{B})$ and $\mathcal{D}_{\Theta_{B}}(k,n_{B})$ related to inhomogeneous magnetic reheating and passive mode, respectively, which determine the amplitude of the angular power spectrum, as a function of $k$ for various values of $n_{B}$.
({\it Bottom}): Dimensionless angular power spectra obtained by Eqs.~(\ref{eq: Cl cross 1}) (dashed) and (\ref{eq: Cl cross 2}) (solid).
In this figure, we set $b_{\rm NL} = 1$, $B_{\lambda} = 1\; {\rm nG}$, $\mathcal{A}_{\zeta} = 2.214\times 10^{-9}$, and $n_{\rm s} = 0.96$.
For comparison, we present plots of the passive mode multiplied by $10^{10}$.
We also show the observed data from Planck~\citep{2018arXiv180706209P} and ACT~\citep{2017JCAP...06..031L}.
Note that we plot the absolute value of $C_{\ell}$.
}
\label{fig: calD and Cl}
\end{figure}

In the case of the power-law type, the power spectrum of PMFs is given as
\begin{align}
\frac{k^{3}}{2\pi^{2}}P_{B}(k) &= \frac{2(2\pi)^{n_{B}+3}}{\Gamma\left( \frac{n_{B}+3}{2}\right)}B^{2}_{\lambda}\left( \frac{k}{k_{\lambda}}\right)^{n_{B}+3} ~,\label{eq: Pmag}
\end{align}
where $k_{\lambda} = 2\pi/\lambda$ and in this paper, we fix $\lambda = 1\; {\rm Mpc}$ as following \citet{2016A&A...594A..19P}.
Note that $B_{\lambda}$ corresponds to the amplitude of the magnetic field after smoothing over the pivot scale $\lambda$.
Then, Eqs.~(\ref{eq: bNL Bzeta zeta}) and (\ref{eq: bNL B zeta}) are rewritten as
\begin{align}
C^{\Theta_{B} \Theta}_{\ell}
&= 
- 6.121\times 10^{-15}
\left( \frac{\mathcal{A}_{\zeta}}{2.214\times 10^{-9}} \right)
\frac{2(2\pi)^{n_{B}+3}}{\Gamma\left( \frac{n_{B}+3}{2}\right)}
\notag \\
&\times 
b_{\rm NL} \left( \frac{B_{\lambda}}{1\; {\rm nG}} \right)^{2}
\int\frac{{\rm d}k}{k}\;
\left( \frac{k}{k_{0}}\right)^{n_{\rm s}-1}
\mathcal{T}^{2}_{\ell}(k)
\mathcal{D}_{\Theta_{B}}(k,n_{B})
\label{eq: Cl cross 1}
~, \\
C^{\Delta_{B} \Theta}_{\ell}
&=
1.991\times 10^{-16}
\left( \frac{\mathcal{A}_{\zeta}}{2.214\times 10^{-9}} \right)
\frac{2(2\pi)^{n_{B}+3}}{\Gamma\left( \frac{n_{B}+3}{2}\right)}
\notag \\
&\times 
b_{\rm NL}\left( \frac{B_{\lambda}}{1\; {\rm nG}} \right)^{2}
\int\frac{{\rm d}k}{k}\;
\left( \frac{k}{k_{0}}\right)^{n_{\rm s}-1}
\mathcal{T}^{2}_{\ell}(k) 
\mathcal{D}_{\Delta_{B}}(k,n_{B}) ~, \label{eq: Cl cross 2}
\end{align}
where we define
\begin{align}
\mathcal{D}_{X}(k,n_{B})
&\equiv 
\int\frac{{\rm d}k_{1}}{k_{1}}\;
\left( \frac{k_{1}}{k_{\lambda}}\right)^{n_{B}+3} \tilde{C}_{X}(k,k_{1}) ~,
\end{align}
with $X = \Delta_{B}$ or $\Theta_B$.
The function $\mathcal{D}_X(k,n_{B})$ generally depends on both $k$ and $n_{B}$ and we show these functions in Fig.~\ref{fig: calD and Cl}.
As we discussed in the previous section, the dominant contribution in the $k$-integral of Eq.~(\ref{eq: Cl cross 2}) is coming from $k\sim \ell/\eta_{0} \ll O(1)\; {\rm Mpc}^{-1}$. 
Then we find the most dominant parts are given as
\begin{align}
\mathcal{D}_{\Theta_{B}}(k\to 0,n_{B})
& \approx
\frac{2^{-(3+n_{B})/2}}{45}
\left( \frac{k}{k_{\lambda}}\right)^{2}
\left( \frac{k_{\rm D}(z_{\nu})}{k_{\lambda}}\right)^{n_{B}+1}\notag \\
&\times 
(n^{2}_{B}+4n_{B}+3)\Gamma\left( \frac{n_{B}+1}{2}\right) ~, \label{eq: calD kto0}\\
\mathcal{D}_{\Delta_{B}}(k\to 0,n_{B})
&\approx
2^{-\frac{n_{B}+3}{2}}
\left( \frac{k_{\rm D}(z_{\rm i})}{k_{\lambda}} \right)^{n_{B}+3}\Gamma\left( \frac{n_{B}+3}{2}\right) ~. \label{eq: approx calD B}
\end{align}
Here, we use the fact that $k_{\rm D}(z_{\rm i}) \gg k_{\rm D}(z_{\nu}) > k_{\rm D}(z_{\mu}) > k_{\rm D}(z_{\rm f})$.
These $\mathcal{D}_{X}$ determine the amplitude of the angular power spectra of the CMB temperature anisotropy.
The function of the passive mode~(\ref{eq: calD kto0}),
is sensitive to the damping scale~$k_D(z_\nu)$
while that of magnetic reheating~(\ref{eq: approx calD B}), is to $k_D(z_i)$.
Therefore, since $k_{\rm D}(z_{\rm i}) \gg k_{\rm D}(z_{\nu})$,
the cross-correlation between $\Theta_B$ and $\Theta$ is strongly suppressed, compared to that between $\Delta_B$ and $\Theta$.
The above features are also found in Fig.~\ref{fig: calD and Cl}, where 
the cross-correlation with the passive modes is plotted with multiplication by $10^{10}$ although that of magnetic reheating is shown without any multiplication.

Finally, the angular power spectra of the CMB temperature anisotropy are given by
\begin{align}
C^{\Theta_{B}\Theta}_{\ell}
&\propto
\int\frac{{\rm d}k}{k}\;
\left( \frac{k}{k_{0}}\right)^{n_{\rm s}-1 + 2}
\mathcal{T}^{2}_{\ell}(k)
~, \label{eq: Cl passive approx} \\
C^{\Delta_{B}\Theta}_{\ell}
&\propto
\int\frac{{\rm d}k}{k}\;
\left( \frac{k}{k_{0}}\right)^{n_{\rm s}-1}
\mathcal{T}^{2}_{\ell}(k)
 ~. \label{eq: Cl cross approx}
\end{align}
From Eqs.~(\ref{eq: Cl passive approx}) and (\ref{eq: Cl cross approx}), the shapes of the angular power spectra induced by non-Gaussian PMFs are the same as the primary CMB spectrum of $n_{\rm s}+2$~(the passive mode) and $n_{\rm s}$~(inhomogeneous magnetic reheating), respectively.
This feature can be shown in Fig.~\ref{fig: calD and Cl}.

In this paper, we set the initial redshift~$z_i$ to that of the electroweak phase transition. However, the correlation
$\Braket{B^{2}\zeta}$ can be initiated when the PMFs are generated during inflation. In this case, the initial
redshift~$z_i$ should go back to reheating epoch, corresponding to the conformal time $\eta_{\nu}/\eta_{B}\approx 10^{17}$.
When we choose this redshift as the initial time, the effect of inhomogeneous magnetic reheating on the CMB temperature anisotropy would be enhanced.
In this sense, we will give conservative results in the next section.

\section{Constraints on PMFs}
\subsection{Upper bounds on Gaussian PMFs}
First, we investigate a constraint on the amplitude of PMFs
from the Gaussian part discussed in Sec.~\ref{sec:Gaussian}.
As shown in the previous section, we found that the disconnected part of $\langle B^4\rangle$
can contribute to $C^{\Delta_{B}\Delta_{B}}_{\ell}$, $C^{\Theta_{B}\Delta_{B}}_{\ell}$, and $C^{\Theta_{B}\Theta_{B}}_{\ell}$.
Then, the angular power spectra of them show the scale dependence similar to those from the blue-tilted adiabatic initial condition with a scalar spectral index $n_s = 3$. Hence, let us take a closer look at
the angular power spectrum on small scales, where the corrections would be relatively enhanced.

The CMB measurement by ACT~\citep{2017JCAP...06..031L} gives the precise data of the small-scale CMB angular power spectrum and the minimum value of the observed amplitude of the CMB temperature fluctuations can be read as $C^{(\rm obs)}_{\ell_{\rm ACT}} \approx 4.3 \times 10^{-18}$ around $\ell_{\rm ACT}
= 3025$, as shown in Fig.~\ref{fig: calD and Cl}.
In the standard cosmology, it is interpreted that
the primordial CMB anisotropy dominates on
scales larger than ACT-scale~($\sim \ell_{\rm ACT}$) while 
the secondary CMB anisotropy including the Sunyaev--Zel'dovich effect of
galaxy clusters and foreground contributions
would dominate the anisotropy on smaller scales.
However, as shown in the previous section,
$C^{\Delta_{B}\Delta_{B}}$ potentially have a significant contribution
on small scales.
Therefore, from the condition $C^{(\rm obs)}_{\ell_{\rm ACT}} >
C^{\Delta_{B}\Delta_{B}}_{\ell_{\rm ACT}}$, inhomogeneous magnetic reheating put an upper bound on PMFs as
\begin{equation}
\mathcal{B} \lesssim 1.3\times 10^{3} \; {\rm nG} ~~~ \mbox{($k_{\rm p} = O(10)\; {\rm Mpc}^{-1}$)} ~.\label{const B disc}
\end{equation}
This constraint is much weaker than that obtained from the CMB
spectral distortion, i.e., ${\cal B} < 10\; {\rm nG}$ on $10\; {\rm Mpc}^{-1} \lesssim k_{\rm p}\lesssim 10^{3}\; {\rm Mpc}^{-1}$~\citep{Jedamzik:1999bm}.
Although $C^{\Theta_{B}\Delta_{B}}_{\ell}$ and $C^{\Theta_{B}\Theta_{B}}_{\ell}$ lead to constraints similar to Eq.~(\ref{const B disc}) on the same scales, that from the CMB spectral distortions are much tighter.
Thus, inhomogeneous magnetic reheating is not useful to constrain Gaussian PMFs.
Indeed, \citet{2015JCAP...05..049N} also showed that the anisotropic
acoustic reheating is not useful to discuss upper bounds on small-scale Gaussian curvature perturbations.

\subsection{Limit on $b_{\rm NL}$}

\begin{figure}
\includegraphics[width=\columnwidth]{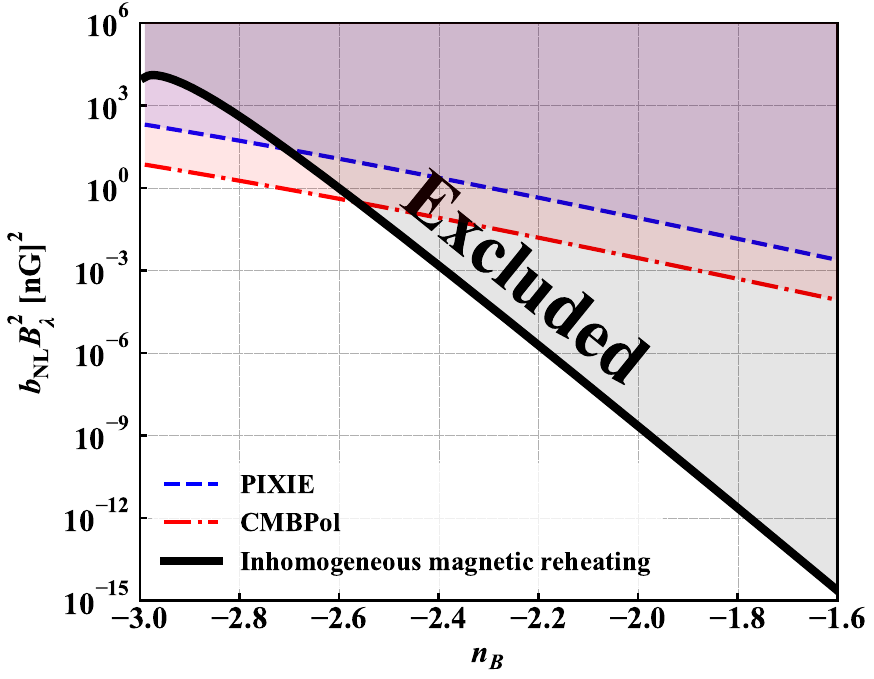}
\caption{%
The upper bound on the combination $b_{\rm NL}B^{2}_{\lambda}$ as the function of $n_{B}$.
The gray-shaded region is excluded by inhomogeneous magnetic reheating.
Here, we use the same model parameters as Fig.~\ref{fig: calD and Cl}.
We also show the forecast from $\Braket{\mu T}$ cross-correlation of Eq.~(6.11) in \citet{Ganc:2014wia}.
}
\label{fig: result cross}
\end{figure}

Let us derive a constraint on the non-linear parameter of
PMFs, which gives mode mixing of large and small scales.
In the previous section, we explored the consequence of the primordial cross-correlation~(\ref{eq: def bNL}).
As seen in the bottom of Fig.~\ref{fig: calD and Cl} and Eq.~(\ref{eq: Cl cross approx}), 
we found that the scale dependence of the angular power spectrum is the same with that of the scale-invariant adiabatic fluctuations, namely, $C^{\Delta_{B}\Theta}_{\ell} \propto C^{\rm Fiducial}_{\ell}$.
The amplitude of $C^{\rm Fiducial}_{\ell}$ seen in Fig.~\ref{fig: calD and Cl} can be fixed by the current observations, and therefore we constrain the signal of inhomogeneous magnetic reheating by $C^{\Delta_{B}\Theta}_{\ell} < C^{\rm Fiducial}_{\ell}$ for an arbitrary $\ell$.
Here, we compared the values of $\ell=2$ to put the constraint on $b_{\rm NL}B^{2}_{\lambda}$, and the result is also shown in Fig.~\ref{fig: result cross}.
In this figure, we also show the upper bound forecasted by using the CMB spectral distortion in Eq.~(6.11) of \citet{Ganc:2014wia}.
It is known that dissipating PMFs can reheat the CMB photons and create $\mu$-distortion during $z_{\rm f}< z < z_{\mu}$.
\citet{Ganc:2014wia} used the cross-correlation between the inhomogeneous $\mu$-distortion and passive mode in order to constrain $b_{\rm NL}$.

In Fig.~\ref{fig: result cross}, we can obtain the approximate form of the upper bound on $b_{\rm NL}B^{2}_{\lambda}$ in $n_{B}$-$b_{\rm NL}B^{2}_{\lambda}$ plane as followings.
By substituting Eq.~(\ref{eq: approx calD B}) into Eq.~(\ref{eq: Cl cross 2}), we obtain
\begin{align}
C^{\Delta_{B}\Theta} &\approx 
b_{\rm NL}\left( \frac{B_{\lambda}}{1\; {\rm nG}} \right)^{2}
2\times 1.991\times 10^{-16}
\left( \frac{k^{2}_{\rm D}(z_{\rm i})}{2}\right)^{\frac{n_{B}+3}{2}} \notag \\
& \times 
\left( \frac{\mathcal{A}_{\zeta}}{2.214\times 10^{-9}} \right)
\int\frac{{\rm d}k}{k}\;
\left( \frac{k}{k_{0}}\right)^{n_{\rm s}-1}
\mathcal{T}^{2}_{\ell}(k) \notag \\
&= 
b_{\rm NL}\left( \frac{B_{\lambda}}{1\; {\rm nG}} \right)^{2}
\frac{2\times 1.991\times 10^{-16}}{2.214\times 10^{-9}}
\left( \frac{k^{2}_{\rm D}(z_{\rm i})}{2}\right)^{\frac{n_{B}+3}{2}}
C^{\rm Fiducial}_{\ell} ~,
\end{align}
where $C^{\rm Fiducial}_{\ell}$ is the CMB temperature anisotropy from the adiabatic curvature perturbation defined by
\begin{equation}
C^{\rm Fiducial}_{\ell} = 
\mathcal{A}_{\zeta}
\int\frac{{\rm d}k}{k}\;
\left( \frac{k}{k_{0}}\right)^{n_{\rm s}-1}
\mathcal{T}^{2}_{\ell}(k) ~.
\end{equation}
Then we impose the condition $C^{\Delta_{B}\Theta}_{\ell} < C^{\rm Fiducial}_{\ell}$ and finally obtain the approximate form on $n_{B}\gg -3$ of the upper limit as $b_{\rm NL}\left( B_{\lambda}/{\rm nG}\right)^{2}$.
In the current setting $z_{i} = 10^{15}$, the upper limit can be given as
\begin{equation}
b_{\rm NL}\left( \frac{B_{\lambda}}{1\; {\rm nG}}\right)^{2}
< {\rm e}^{-36.5n_{B} - 94.0} ~.
\end{equation}
We can find this approximate form in Fig.~\ref{fig: result cross} in the case of large $n_{B}$.

Although the upper bound from inhomogeneous magnetic reheating is coming from the current Planck observation, it can put a tighter constraint on the primordial non-Gaussianity in PMFs than the forecasts based on the CMB spectral distortion in \citet{Ganc:2014wia}, especially in the case of the bluer tilted magnetic fields.
This is the advantage of inhomogeneous magnetic reheating which can be sensitive to smaller scales as shown in Fig.~\ref{fig: C k}.

\section{Summary}

In this paper, we explore additional CMB temperature anisotropies from the dissipation of PMFs and derive some constraints on the statistical quantities of PMFs.

Fast magnetosonic modes originated from PMFs dissipate on small scales so that the energy of PMFs in a local comoving volume is decreasing.
Thanks to the energy conservation laws of the total system of radiation and magnetic fields, the released energy would be transferred to radiation.
If PMFs are initially fluctuating, the energy injection into radiation should be also fluctuating, which leads to additional temperature perturbations.
We call such a secondary CMB temperature anisotropy \textit{inhomogeneous magnetic reheating}, which has never been considered in the literature.
Inhomogeneous magnetic reheating should be subdominant part of the
CMB anisotropy; therefore, we can put some upper bounds on the
statistical properties of PMFs, imposing inhomogeneous magnetic reheating not to exceed the observed CMB temperature power spectrum.

First of all, we give a formulation of the secondary CMB temperature anisotropy from inhomogeneous magnetic reheating as a straightforward extension of global magnetic reheating proposed by~\citet{Saga:2017wwr}.
Then, we evaluate the corrections to the CMB temperature power spectrum that originated from Gaussian PMFs.
We find that inhomogeneous magnetic reheating can have relatively huge contributions on the scales where the passive mode is not produced: i.e., $10^{5}\; {\rm Mpc}^{-1}\lesssim k$.
However, the final expression of the secondary temperature anisotropies is highly suppressed on the CMB anisotropy scales so that we cannot get a meaningful constraint.
This is because the secondary temperature power spectrum from disconnected four-point PMFs is mainly produced on the scales where magnetic reheating actually happened, which are far smaller than the damping feature in the temperature angular power spectrum.
As we see in Fig.~\ref{fig: calD and Cl}, we observe the low $\ell$ tail of the blue-tilted power spectrum, which is suppressed on the CMB scales.

On the other hand, we find that inhomogeneous magnetic reheating has the advantage to probe non-Gaussian PMFs on small scales, which can be another hint of primordial magnetogenesis.
Here, we assume local-type non-Gaussianity of PMFs parametrized in Eq.~(\ref{eq: def bNL}).
This shape of non-Gaussianity leads to the mode mixing between small-scale magnetic fields and a large-scale curvature perturbation.
Hence, inhomogeneous magnetic reheating creates additional long-wavelength temperature perturbations which are correlated with original long-wavelength curvature perturbations.
In this case, we obtain the strongest upper limit on the amplitude of local-type PMF non-Gaussianity.
Although we mainly focus on the phenomenological parametrization throughout this paper, exploring the explicit model of primordial non-Gaussianity in PMFs will be left as future work.

\section*{Acknowledgements}
This work is supported by a Grant-in-Aid for Japan Society for Promotion of Science (JSPS) Research Fellow Number 17J10553~(SS), JSPS Overseas Research Fellowships~(AO), JSPS KAKENHI Grant Number 15K17646~(HT) and 17H01110~(HT), and MEXT KAKENHI Grant Number 15H05888~(SY) and 18H04356~(SY).
\bibliographystyle{mnras}
\bibliography{ref}
\bsp 
\label{lastpage}
\end{document}